\RequirePackage{amsmath}
\documentclass[runningheads]{llncs}
\usepackage{amsmath,amsfonts}
\usepackage{algorithmic}
\usepackage{algorithm}
\usepackage{array}
\usepackage[caption=false,font=normalsize,labelfont=sf,textfont=sf]{subfig}
\usepackage{textcomp}
\usepackage{stfloats}
\usepackage{url}
\usepackage{verbatim}
\usepackage{graphicx}
\usepackage{cite}
\usepackage{amssymb}
\usepackage{enumerate}
\usepackage{subfig}
\usepackage{setspace}
\usepackage{bm}
\usepackage{textcomp}
\usepackage{xcolor}
\usepackage[scr=boondoxo,scrscaled=1.05]{mathalfa}
\usepackage{upgreek}
\usepackage{marginnote}

\usepackage{hyperref}

\newcommand{\be}{\begin{equation}}
\newcommand{\ee}{\end{equation}}
\newcommand{\bee}{\begin{equation*}}
\newcommand{\eee}{\end{equation*}}
\newcommand{\ba}{\begin{array}}
\newcommand{\ea}{\end{array}}
\newcommand{\baa}{\left[\begin{array}}
\newcommand{\eaa}{\end{array}\right]}

\newcommand{\beqa}{\begin{eqnarray}}
\newcommand{\eeqa}{\end{eqnarray}}
\newcommand{\bt}{\begin{tabular}}
\newcommand{\et}{\end{tabular}}

\newcommand{\bi}{\begin{itemize}}
\newcommand{\ei}{\end{itemize}}
\newcommand{\bc}{\begin{center}}
\newcommand{\ec}{\end{center}}

\newtheorem{Defi}{Definition}

\newlength{\algitab}
\newcommand{\bbm}{\begin{bmatrix}}
\newcommand{\ebm}{\end{bmatrix}}

\begin{document}

\title{Host Attack Graph for Botnet Propagation}



\author{
  Andrei Neagu\textsuperscript{$\star$}
  \and
  Mara-Cristina Sterian\textsuperscript{$\star$}
  \and
  Paul Irofti\textsuperscript{$\dagger$}
}

\authorrunning{Andrei Neagu, Mara-Cristina Sterian, Paul Irofti}

\institute{
  Department of Computer Science,
  Faculty of Mathematics and Computer Science,
  and Interdisciplinary School for Doctoral Studies, 
  University of Bucharest, Romania\\
  \email{andrei@neagu.email, mara-cristina.sterian@unibuc.ro, paul@irofti.net}\\
  \textsuperscript{$\star$}These authors contributed equally to this work. \quad
  \textsuperscript{$\dagger$}Corresponding author.
}

\date{}

\maketitle

\begin{abstract}
Botnets represent the backbone of most modern cybersecurity attacks
through which botmasters deploy distributed denial of service
and advanced persistent threats
against legitimate computer networks.
In this paper we introduce the Host Attack Graph model
and propose two botnet propagation strategies
which we study
across network topologies,
target selection,
and attack strategies
together with their effectiveness in time.
While strategy and botnet size
are important for propagation,
our study and simulations show that
network topology and target selection
should not be easily discarded.
Implementation and benchmarks are available at 
\url{https://codeberg.org/AndreiN/HAGP}.

\end{abstract}
\section{Introduction}

Botnets are computer networks comprised of compromised machines (or bots)
and rely on the ability of the attacker (or botmaster) to coordinate the victims~\cite{Mahboubi-IOT-2025-EvolvingBotnets}
to employ various types of attacks
such as distributed denial of service (DDoS)~\cite{Us-AAA-2026-ddos,Hoque-ICST-2015-botddos},
data exfiltration~\cite{Nguyen-DGTS-2017-Exfiltration,Venkatesan-SP-2018-debot},
phishing and spamming~\cite{Xie-SIGCOMM-2008-spamming,Khan-ICST-2015-SpamSurvey},
advanced persistent threats~\cite{Chen-ACM-2022-SurveyAPT},
but also bot network expansion (or propagation)~\cite{Vormayr-ICST-2017-botprop,Xia-IOTJ-2020-socialbotprop}.
Recent takedowns showed networks as large as
19 million bots~\footnote{\url{https://www.bloomberg.com/news/articles/2024-05-29/fbi-takes-down-massive-global-army-of-zombie-computer-devices}}
with the DDoS defence market estimated to reach 7.3 billion USD in 2027~\footnote{\url{https://www.marketsandmarkets.com/Market-Reports/ddos-protection-mitigation-market-111952874.html}}.

Botmasters control botnets through either a centralised command and control~(C2) server,
a decentralised topology in which commands are relayed across
adjacent nodes, or a hybrid of both.
A given botnet is described through its
structural features such as
propagation, registration of newly compromised bots (or rallying),
establishing (alternative) communication channels
and evading detection~\cite{Dagon-ACSAC-2007-BotnetTaxonomy}.
Attackers issue commands and collect harvested data
through one or more C2 servers,
through which they coordinate the bots.
Being crucial to the botnet and the attack,
the C2 server is carefully concealed and protected 
by the malicious actor.
After the initial infection,
the newly acquired bot attempts to contact
the C2 server,
typically via an embedded domain name,
or one of its peers
to announce its presence and
receive further instructions.
This process is typically referred to
as \textit{call-to-home} or \textit{rallying}.

Botnet capabilities are influenced by a series of factors 
such as the network structure, bandwidth, robustness and communication structure.
Nodes occupying different structural positions
and
contribute unequally to the spread of an infection.
A central, highly connected host 
can serve as an influential attack vector,
as opposed to a peripheral host.
This structural heterogeneity is formalised 
through centrality measures, such as
degree, closeness, betweenness, and eigenvector centrality.
A node's relative importance can be computed
according to a node's connectivity, 
proximity to other nodes, 
or influence within the graph.
Therefore, finding these high-scoring nodes
represents a very important step 
in characterizing malware propagation behaviour
and designing targeted defensive strategies.

\textbf{Our contributions} in this paper are:
    1) botnet attack modelling: we introduce Host Attack Graph (HAG), a novel method of computing root compromise probabilities,
    along with the Susceptible, Infected, Protected (SIP) model for the network evolution across time
    and two new strategies for target selection and propagation;
    2) a study and Spearman analysis across different graph topologies, node centralities and propagation strategies
    where we introduce for comparison purposes
    a new ideal strategy for root greedy propagation 
    and a baseline random choice algorithm;
    besides multiple scenarios across varied parameter choices and randomization,
    our study and simulations also include a Spearman analysis across centralities;
    3) we make available the source code and datasets used to compare botnet propagation algorithms
    including probability computations, centrality and scenario generation.

\textbf{Outline.}
The rest of the paper is structured as follows.
In Section~\ref{sec:soa} we introduce the state of the art,
followed by our host attack graph, network and propagation models in Section~\ref{sec:hag}
and the experiments from Section~\ref{sec:sim}.
We conclude and provide future directions in Section~\ref{sec:end}.

\section{State of the art}\label{sec:soa}

A taxonomy of botnet structures is proposed in
\cite{Dagon-ACSAC-2007-BotnetTaxonomy}.
The authors identify the three major measures of botnets: 
robustness, effectiveness, and efficiency,
then establish three graph models to assess the first: 
the Erdős–Rényi Random Graph Model, 
the Watts-Strogatz Small World Models, 
and the Barabási-Albert Scale Free Models.
In this paper, 
we use these graph model families 
to test the botnet's propagation capabilities
under different spreading strategies.

Regarding botnet propagation,
the authors in \cite{Yu-TKDE-2015-PropagLarge}
study propagation in large-scale networks
with a two-layer SI epidemic model approach: 
an upper layer (the Internet level)
and
a lower layer (the network level).
All the factors that influence the spreading of the botnet 
are reduced to the infection rate $\beta$ 
which is applied to the
Android malware \cite{Zhou-SSP-2012-Android} and Conficker \cite{Shin-TIFS-2012-Conficker} datasets,
in order to test their theoretical findings.
Their proposed model predicts that 
malware propagation across networks
passes through three stages:
exponential distribution, 
power law distribution with a short exponential tail 
and power law distribution.
In our work,
instead of using a single scalar infection coefficient,
we assign to each host 
its own probability of compromise, 
computed from
a set of remote and local vulnerabilities,
quantified by their CVSS scores 
and chained into attack paths.

In order to model the propagation process in time,
the authors in 
\cite{Martin-Neurocomputing-2021-MalwareProp}
employ cellular automata 
with a set of four device states 
(susceptible, infected, attacked and recovered) 
to highlight the dynamics of the network system
and
analyse malware propagation
on Erdős-Rényi random networks
by choosing, 
as their transmission vectors, 
the nodes with highest centrality measures
(degree centrality, betweenness centrality, closeness centrality, eigenvector centrality).
Their findings order centrality measures 
by infection efficiency, 
from the most to least efficient, 
as follows: 
eigenvector centrality, closeness centrality, betweenness centrality, degree centrality.
We integrate these as spreading strategies in our work.
In \cite{Oldham-PONE-2019-Spearman} the authors
studied the correlation between $17$ centrality measures across
$212$ real-networks.
Their research is 
computing the Spearman matrix
and conducting a principal component analysis (PCA).
We note here that the Spearman correlation 
is utilised to measure the strength of the monotonic
relation between two ranked variables.
Thus, $1$ represents a perfectly increasing relationship,
while $-1$ a perfectly decreasing one.
The resulting correlations were mostly positive
as $97\%$ of them exceeded the
$0.5$ threshold in unweighted networks.
PCA showed that centrality is largely unidimensional
in most networks.

In \cite{Poolsappasit-TDSC-2012-BAG}
a novel approach for graph modelling is proposed
based on Bayesian belief networks called
Bayesian Attack Graphs~(BAG).
Their work incorporates different causality relationships
between network states.
In our work,
we expand on their proposed method, at the host level,
in order to determine the likelihood that a host is compromised.
A discrete-time SIR (Reed-Frost model) analysis is presented in 
\cite{Draief-AAP-2008-Thresholds}. 
At every time slot $t$, 
every infected node attempts to infect each of its neighbours 
with an independent success probability $\beta$.
If the infection is successful the node is removed.
They find that the number of infected hosts 
is directly dependent on the network topology.
The model implies
a single homogenous infection probability across all edges 
and permanent removal, 
with no vulnerability structure or attacker persistence.
We consider these aspects important for botnet analysis
and include them in our proposed model.

\section{Modelling the Attack}\label{sec:hag}
In this section we present
the construction of the Host Attack Graph (HAG) model
together with the Susceptible Infected Protected (SIP) network evolution model 
and the calculation of centralities used in target selection for propagation.

\subsection{Host Attack Graph}
Host Attack Graph (HAG) is built 
on the framework of a Bayesian Attack Graph~\cite{Poolsappasit-TDSC-2012-BAG},
defined at the individual machine level.
Attributes represent the host's vulnerabilities, denoted as $v$,
alongside attacker goals $U(h)$ and $R(h)$
representing the user and root privileges that the attacker can obtain.
The Common Vulnerability Scoring System \cite{cvss-manual},
$CVSS(v)$, represents the severity score of a
vulnerability
denoted as an integer between $[0, 10]$,
where 10 ($CVSS_{MAX}$) represents the highest severity
score a vulnerability can take.
Vulnerability $v$ has an associated probability of successfully being exploited by an attacker $p_e(v)$.
In our paper we use a synthetic value for this based on its known severity score $\frac{CVSS(v)}{CVSS_{MAX}}$.
While this is not ideal as the two measures are not linearly connected,
we argue that this is sufficient for our simulations
and consider the study of $p_e(v)$ modelling outside the scope of this paper.

\begin{figure}[t]
    \centering
    \includegraphics[width=1\textwidth]{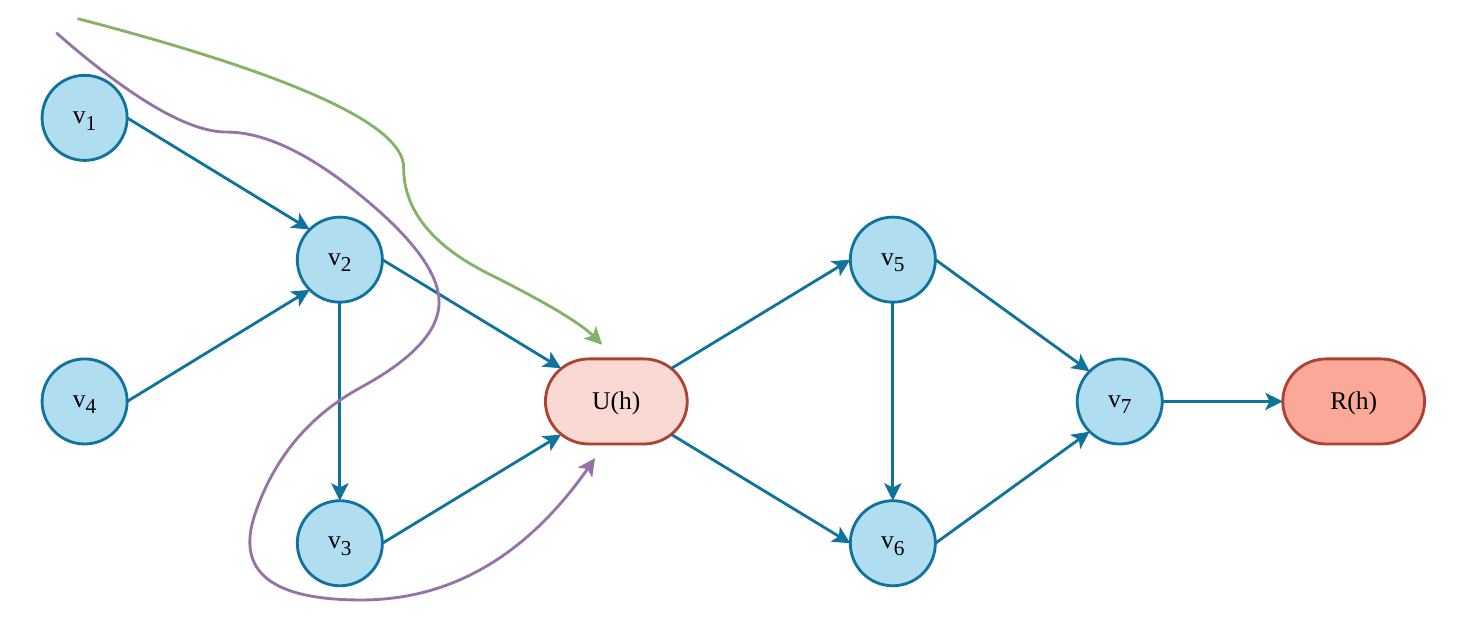}
     \caption{Example of a Host Attack Graph (HAG).
     When chained together, host's $h$ vulnerabilities $v_1$,  $v_2$, $v_3$, $v_4$ generate attack paths
     that are used to obtain user privileges $U(h)$;
     the green and purple lines represent valid attack paths.
     Once $U(h)$ is achieved,
     vulnerabilities $v_5$, $v_6$, $v_7$ are used to form attack paths towards root access.}
    \label{fig:hag}
\end{figure}

\begin{Defi}[Vulnerability Attributes]
    Given host $h$ we denote its set of vulnerabilities with $V(h) = \{v_1, \dots v_n\}$.
    Let $S_v$ with $v \in V$ represent the Bernoulli trials with probability $p_e(v)$.
    Then the vulnerability attributes of $h$ is the set
    $A_v(h) = \{ S_v: v \in V(h)\}$,
    where $S_v = 1$ denotes that $v$ has been successfully compromised with $Pr(S_v = 1) = p_e(v)$.
\end{Defi}

\begin{Defi}[Goal Attributes]
    A goal $g$ on host $h$ is represented by a vulnerability subset $V(g(h))$.
    We say that goal $g$ is reached if and only if there exists a vulnerability $v \in V(g(h))$ such that $S_v = 1$
    (i.e.~vulnerability $v$ was successfully exploited by the attacker to reach goal $g$).
    We denote the set of goal attributes on host $h$
    with $G(h) = \{ g_1, g_2, \dots, g_n \}$
    such that $g_i = V(g_i(h)) \subseteq V(h),\ \forall i$.
\end{Defi}

\begin{remark}
    In practice,
    the attacker has to first exploit a chain of vulnerabilities
    in order reach and exploit $v \in V(g(h))$.
    These are defined later and not included in $V(g(h))$.
\end{remark}
\begin{remark}
    Of special interest in this paper,
    for a given host $h$,
    are the goals of user and root privileges respectively
    for which we have the separate notation $U(h)$ and $R(h)$.
\end{remark}

\begin{example}
    In Figure~\ref{fig:hag} we have goal attributes $U(h)$ and $R(h)$
    with vulnerability subsets $g_{U} = \{ v_2, v_3 \}$ and $g_{R} = \{ v_7 \}$.
\end{example}

\begin{Defi}[Host Attributes]
    Given host $h$ with vulnerability attributes $A_v(h)$ and goal attributes $G(h)$
    the host attributes are $A(h) = A_v(h) \cup G(h)$.
\end{Defi}

\begin{example}
The attributes of host $h$ from Figure~\ref{fig:hag} are
$A(h) = \{ S_v: v \in \{v_1 \dots v_7\} \cup \{ U(h), R(h)\}$.
\end{example}

%

\begin{remark}
    Please note that on any given host $h$,
    before the attacker can exploit $v \in V(g(h))$, in order to attain $g(h)$,
    it might be necessary to first exploit other vulnerabilities
    that open-up the opportunity to attempt $v$.
\end{remark}

\begin{Defi}[Vulnerability prerequisite]
    Given vulnerabilities $v$ and $w$ such that
    $v$ can only be reached when $S_w = 1$,
    we say that $w$ is a prerequisite vulnerability of $v$
    and denote this relationship by $w \sqsubseteq v$.
\end{Defi}

\begin{example}
    In order to reach $U(h)$ in Figure~\ref{fig:hag} via $v_2$,
    either $v_1$ or $v_4$ have to be exploited first,
    thus $v_1 \sqsubseteq v_2$ and $v_4 \sqsubseteq v_2$.
\end{example}

\begin{Defi}[Vulnerability chain]
    We say that vulnerabilities $v_1, v_2, \dots, v_n$ form the prerequisite chain
    $v_1 \sqsubseteq v_2 \sqsubseteq \dots \sqsubseteq v_n$
    if and only if
    $v_1$ has no other prerequisites
    and $v_{i-1}$ is the vulnerability prerequisite of $v_{i}$,
    $\forall i = \overline{2,n}$.    
\end{Defi}

\begin{Defi}[Vulnerability Partition]
    Given $V(h)$,
    the set of vulnerabilities
    of host $h$,
    let $V_U(h)$ and $V_R(h)$ be the subsets which contain
    the vulnerabilities of all possible chains
    that may grant the attacker privileges to the host
    user and the root, respectively,
    such that
    \be
        V_U(h) \cup V_R(h) = V(h)\ ;\ \; \ 
        V_U(h) \cap V_R(h) = \emptyset
        \label{eq:v-set-independent}
    \ee
\end{Defi}

\begin{example}
The vulnerabilities from Figure~\ref{fig:hag}
are partitioned into 
$V_U(h) = \{ v_1, v_2, v_3, v_4 \} $ and $V_R(h) = \{ v_5, v_6, v_7 \}$.
\end{example}


\begin{Defi}[Attack Path]
\label{def:attack-path}
    Let $g(h) \in G(h)$ be a goal attribute of host $h$.
    We say that a vulnerability chain is an attack path $\pi(g(h))$
    if its joint vulnerability exploitation is sufficient to gain $g(h)$
    (i.e.~for all $v$ in the chain $S_v = 1$).
    We denote the set of all attack paths leading to $g(h)$ by
    $\Pi(g(h)) = \{ \pi_1(g(h)), \dots, \pi_k(g(h)) \}$.
\end{Defi}

\begin{remark}
\label{remark:goal-attributes}
    The goal attributes can be expressed as
    OR-decomposed over the attack paths and
    AND-decomposed over the vulnerabilities in each path:
    $U(h) = \bigvee_{\pi_i \in \Pi(U(h))} \bigwedge_{v \in \pi_i}S_v$,
    $R(h) = U(h) \wedge \bigvee_{\pi_i \in \Pi(R(h))} \bigwedge_{v \in \pi_i}S_v$.
\end{remark}

\begin{Defi}[Probability of User Compromise]
    The probability that an attacker can successfully gain user access
    $Pr(U(h)=1)$ can be written as the
    probabilistic principle of inclusion and exclusion (PPIE) over
    dependent events
   $Pr(U(h)=1) = \sum_{i=1}^{k}(-1)^{i + 1} \left(
   \sum_{1 \leq i_1 < \dots < i_k \leq n} Pr(\pi_{i_1} \cap \dots \cap \pi_{i_k} = 1)
   \right)$.
\end{Defi}

\begin{Defi}[Probability of Root Compromise] \label{def:pr_def}
    The degree to which an attacker can successfully
    obtain $R(h)$ is dependent of the host level access
    being obtained.
    However \eqref{eq:v-set-independent} suggests
    that the attack paths sets of each goal attribute are independent.
    Thus, the probability of root compromise can be expressed
    as the product between $Pr(U(h)=1)$ and 
    the PPIE over local attack paths.
    \be
        Pr(R(h) = 1) = Pr(U(h)=1) \sum_{i=1}^{k}(-1)^{i + 1} \left(
        \sum_{1 \leq i_1 < \dots < i_k \leq n} Pr(\pi_{i_1} \cap \dots \cap \pi_{i_k} = 1)
        \right)
    \label{pr_def}
    \ee
\end{Defi}

\begin{Defi}[Host Attack Graph]
    The host attack graph of host $h$ is the tuple
    $HAG(h) = (V(h), A_v(h), G(h), \Pi(R(h)))$
    where
    $V$ is the set of vulnerabilities,
    $A_v$ the set of vulnerability attributes,
    $G$ the set of goals,
    and $\Pi$ the set of all attack paths
    leading to the probability of root compromise on the host.
\end{Defi}

\subsection{Network Model}\label{Network_Topology}

To measure behavioural patterns of propagation,
we consider three types of graphs 
studied in the extensive literature on complex networks.
First,
the Erdős-Rényi random graphs~\cite{Erdos-1960-ER}
connect each pair of hosts
independently, resulting in a binomial degree distribution
centred around its mean. 
These networks are structurally uniform, 
thus no selection heuristic can benefit 
from differences in node degree.
We will utilise this model as a baseline for our comparison.
Next,
Barabási-Albert scale-free model~\cite{Barabasi-MP-2002-ScaleFree}
builds the graph by preferential
attachment,
thus creating a power-law distribution
where few hubs have much more connectivity than others.
The authors in \cite{Faloutsos-1999-SIGCOMM-PowerLaw}
suggest that the Internet follows this distribution at both the
domain-level and the router-level.
These scale-free topologies are found in service-dependable
organisations where a small set of infrastructure hosts
are accessed by nearly every workstation.
Lastly,
Watts-Strogatz small-world graphs~\cite{Watts_Strogatz-Nature-1998-SmallWorld}
share similarities with enterprise networks
as tightly connected subnets 
are linked by a few long-range connections.
Here, local clustering is high,
while shortcuts reduce the typical path length.

Regardless of its topology,
the network is also
an undirected graph $G = (V, E)$,
where each vertex represents a host 
and each edge indicates a connection between two hosts.
Hosts in the network are divided into three types:
hosts exposed to the exterior of the network
that are susceptible to be taken over by the attackers,
denoted as $S$,
hosts that are under the attacker's control $I$,
and hosts that only communicate within the network
and are thus protected by the attacker
denoted as $P$,
Accordingly, each node can be in one of three possible states: 
susceptible (S), infected (I) and protected (P) making this a SIP model.

\begin{figure}[t]
    \centering
    \includegraphics[width=1\textwidth]{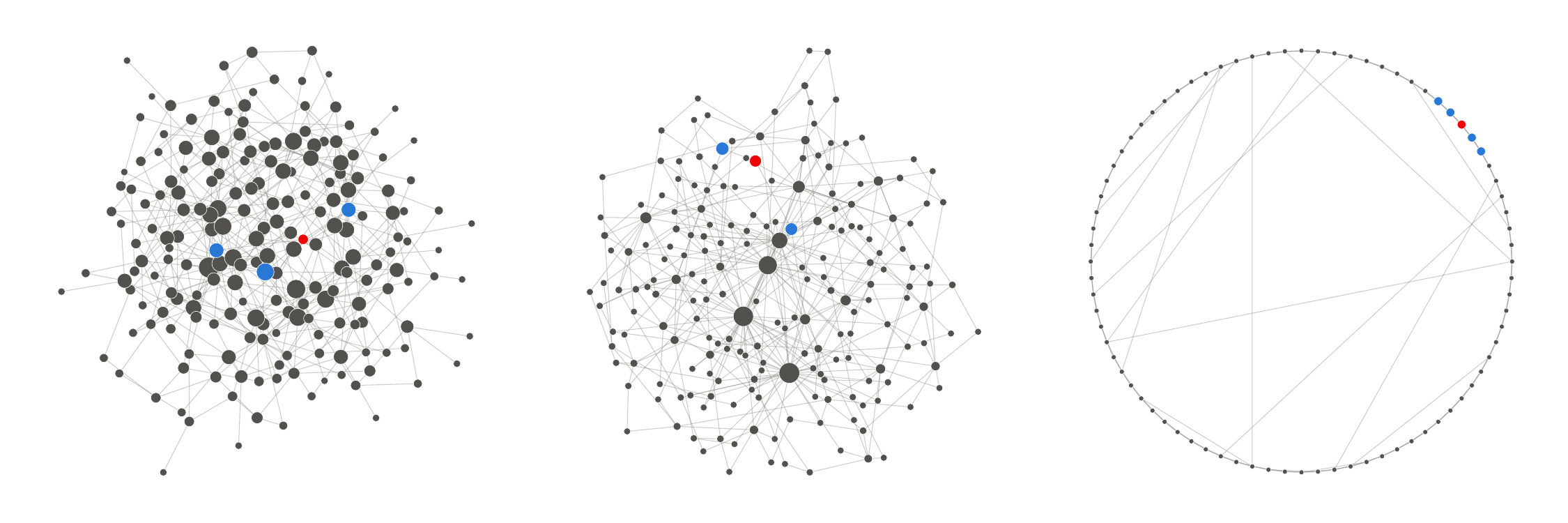}
    \caption{Example graphs ($S$ - blue, $I$ - red, $P$ - grey).
    From left to right: a random graph ($n = 200$), a scale-free
    graph ($n = 200$) and a small-world graph ($n = 80$).
    Susceptible ($S$) and infected ($I$) nodes are artificially inflated
    for visibility,
    otherwise size represents the degree of connectivity in the network.}
    \label{fig:graph_topologies}
\end{figure}

Further on we analyse this model and its evolution across time.
Keep in mind that at time $t$ in our model
an attacker cannot directly compromise
a host that is not directly linked to:
the nodes in $I^{(t)}$ can only attack the ones in $S^{(t)}$.
In Figure \ref{fig:graph_topologies}
we depict the three types of network topologies used throughout the paper
at a fixed moment in time
where the $S$ (blue), $I$ (red) and $P$ (grey) sets are marked accordingly;
we choose small network sizes $n$ as it is easier to visualize and distinguish them this way.

In the following we define
    the frontier as the set of susceptible hosts at time $t$.
    Assuming $I^{(t)}$ is known,
    the susceptible hosts are the ones adjacent to at least one infected node
    $S^{(t)} = \{ s \mid i \in I^{(t)}, h \in V \setminus I^{(t)} \; \text{s.t.} \;  (s,h) \in E \}$.
    It follows that the protected set is $P^{(t)} = V \setminus I^{(t)} \setminus S^{(t)}$.

\subsection{Propagation}\label{sec:target_selection}

The selection pool is always the frontier, the hosts in $S^{(t)}$. 
A host that does not share an edge with the attacker is never a target, 
no matter how important or central it is to the network
as it is unreachable by the attacker.
The attacker selects $\min(k, |S^{(t)}|)$ hosts to compromise 
during each interval 
and performs a single attack (rolls once) for each of them 
against their $Pr(R)$,
from \eqref{pr_def}.
Target selection is realised before any roll is made, 
so if a roll is successful and a host is compromised at $t$,
the frontier will grow at $t+1$, not at $t$.
As a result, an early win does not change 
the target number for that interval.
Given that 
the probability of compromise
remains hidden from the attacker
at any point in time,
target selection is based on centrality measures.
In our paper
we focused on the 
most commonly used centralities
from the network analysis literature
such as
degree,
closeness,
betweenness,
eigenvector
and
percolation
based centralities
(see also Appendix Table~\ref{tab:centralities}).
Next, we proceed to discuss each model.

Degree centrality $d_v$ refers to the number of edges incident to a node.
In this work we use the normalised variant,
so degree centrality of node $v$
is the fraction between 
the node's degree  $\deg(v)$
and the total number of edges $m$ in the graph
such that $d_v = \frac{\deg(v)}{m}$.
Closeness centrality $c_v$ measures how close a node is 
to the other nodes in the graph, 
and is defined as 
the reciprocal of the average 
shortest path distance $d(u,v)$
to the node over 
all the other $n-1$ nodes in the graph reachable by node $v$:
$c_v = \frac{n-1}{\sum_{u=1}^{n-1} d(u,v)}$.
Betweenness centrality $b_v$ is measured 
as the sum, 
over all the pair of nodes $(s,t)$, 
in the graph $V$
of the ratio
$\sigma(s,t \mid v) / \sigma(s,t)$, 
where $\sigma(s,t)$ is the number of shortest paths between $s$ and $t$,
and $\sigma(s,t \mid v)$ is the number of shortest paths between $s$ and $t$ passing through $v$:
$b_v = \sum_{s,t \in V} \frac{\sigma(s,t \mid v)}{\sigma(s,t)}$.
Eigenvector centrality implies that 
links to high-scoring nodes matter more 
than links to lower-scoring nodes.
Thus, the eigenvector centrality value
of a node is proportional to the sum
of the centrality values of the node's predecessors $N(v)$:
$e_v = \frac{1}{\lambda} \sum_{u \in N(v)} e_u$.
Percolation centrality~\cite{Piraveenan-PLoS-2013-Percolation}
for a node is defined as the proportion of shortest paths
between a pair of nodes, in which the source node is
percolated ($x_i = 1)$.
Each node $i$ carries a state $x_i \in [0,1]$
representing the extent to which it is percolated
at time $t$.
The authors define the centrality of a node $v$ as
$p^{(t)}_v = \frac{1}{n-2}\sum_{s \neq v \neq t} \frac{\sigma(s,t \mid v)}{\sigma(s,t)} \frac{x^{(t)}_s}{[\sum x^{(t)}_i]-x^{(t)}_v}$,
where $\sigma(s,t), \sigma(s,t \mid v)$ are the shortest
paths defined from betweenness.
When every node carries the same state,
percolation is structurally the same as betweenness centrality.
Since centralities are evaluated once,
at the moment the attacker enters the network,
we fix the percolation values across the attack time horizon
and chose the binary states: 
$x_i = 1$ for each infected node $i \in I$ 
and $x_j = 0$ for the remaining nodes $j \in V \setminus I$.
In our work,
we assume that the attacker has already
carried out a successful reconnaissance attack
of the entire network topology.
Thus,
the centralities are computed once,
at the start of the run,
and are not updated as the attack progresses.
This can be a short-coming for long-term attacks
which we plan to analyse in the future.

Finally,
we propose two attack strategies that can be paired with
each centrality.
The \textit{sticky} strategy selects the top-$k$ ($\arg\max_k$) susceptible hosts
with the highest centrality and tries to compromise them 
at each time $t$.
Let $c :  V \to \mathbb{R}$ be the centrality score for any given centrality
such that $c(h_1) \leq c(h_2) \leq \dots c(h_n)$ is ordered.
We introduce the \emph{attacker budget} $k \in \mathbb {N}$,
which models how many hosts can be targeted during each time window by the botnet.
Formally the sticky strategy becomes
$\text{sticky}^{(t)}_k = \arg\max_k(c, S^{(t)})$.
The \textit{weighted} strategy selects $k$ susceptible hosts
at random, weighted by their given centrality score,
at each interval $t$.
As we will see in the next section,
every strategy and centrality paring 
is compared against two baselines,
\textit{random sticky} and \textit{random weighted}.
The former draws a host at random and keeps attacking it
until it is compromised,
while the latter selects a host randomly at each interval $t$.
Furthermore, 
we include an \textit{ideal} attacker,
which always targets the host with the highest probability
of being compromised,
as a reference point.

    



\section{Experiments}\label{sec:sim}

Experiments were conducted on a
AMD Ryzen 9 9900X 12-Core processor
with 64 GB of RAM
under Linux 6.8.0,
using Python 3.12.3 with NetworkX 3.6.1, NumPy 2.5.1, and SciPy 1.18.0\footnote{Our code and data is available online at \url{https://codeberg.org/AndreiN/HAGP}}.
For the HAG model from Section~\ref{sec:hag},
vulnerabilities, attack paths and initial states
are randomly assigned for each host.
Each node gets between $[1,3]$ user and root vulnerabilities
with uniformly drawn CVSS scores used as $p_e(v)$.
The number of attack paths are randomly chosen between $[1,3]$,
each with a random user and root vulnerabilities subset.
Trials redraw the attacker chance against $p_e(v)$ 
keeping the HAG and network topology constant.
There is no correlation between a node’s graph position
and the vulnerabilities it uses.
For comparison,
the graph topologies from Section~\ref{Network_Topology}
are generated with the same number of nodes $n$ and edges $m$.
In the case of small-world graphs, 
$m$ represents the target number of edges,
as the ring lattice degree $n_k$
is derived from the target average degree $\frac{2m}{n}$.
Scale-free graphs start from 
an initial star graph of size $L+1$, with $L$ edges, 
and grow by adding nodes up until $n$.
Each new node links to $L$ already existing nodes.
Preferential attachment ensures that 
the more connected a node is,
the more likely is to receive a new link.
To ensure the final number of edges $m$ stays
roughly the same as for the other two families,
we use $L$ = $\min(\max(1, \frac{m}{n}), {n-1})$.
As size and density remain almost the same, 
only the structure differs
such that
differences in propagation 
are attributable to topology.
Successfully exploiting a host
gives the attacker access to its frontier (see Section~\ref{sec:target_selection}).
All experiments include sticky and weighted strategies
compared against 
random sticky and random weighted baselines,
and an ideal strategy.

\begin{figure}[t]
    \centering
    \includegraphics[width=1\textwidth]{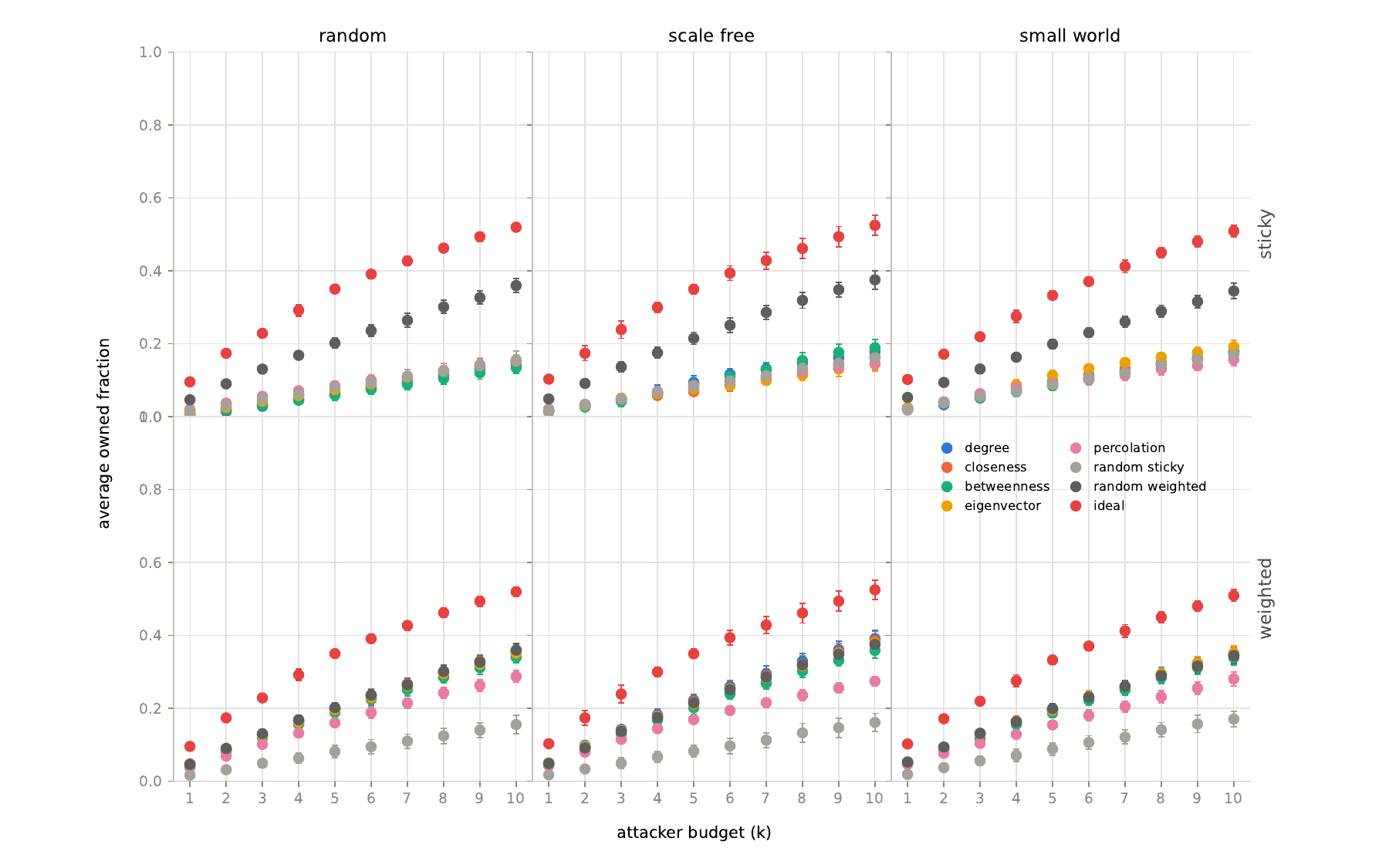}
     \caption{Percentage of owned hosts as the attacker budget $k$ is increased. The rows represent propagation strategies and columns graph topologies. Data points are means over 100 trials and bars represent the standard deviation.}
    \label{fig:attack_scaling}
\end{figure}

\begin{figure}[t]
    \centering
    \includegraphics[width=1\textwidth]{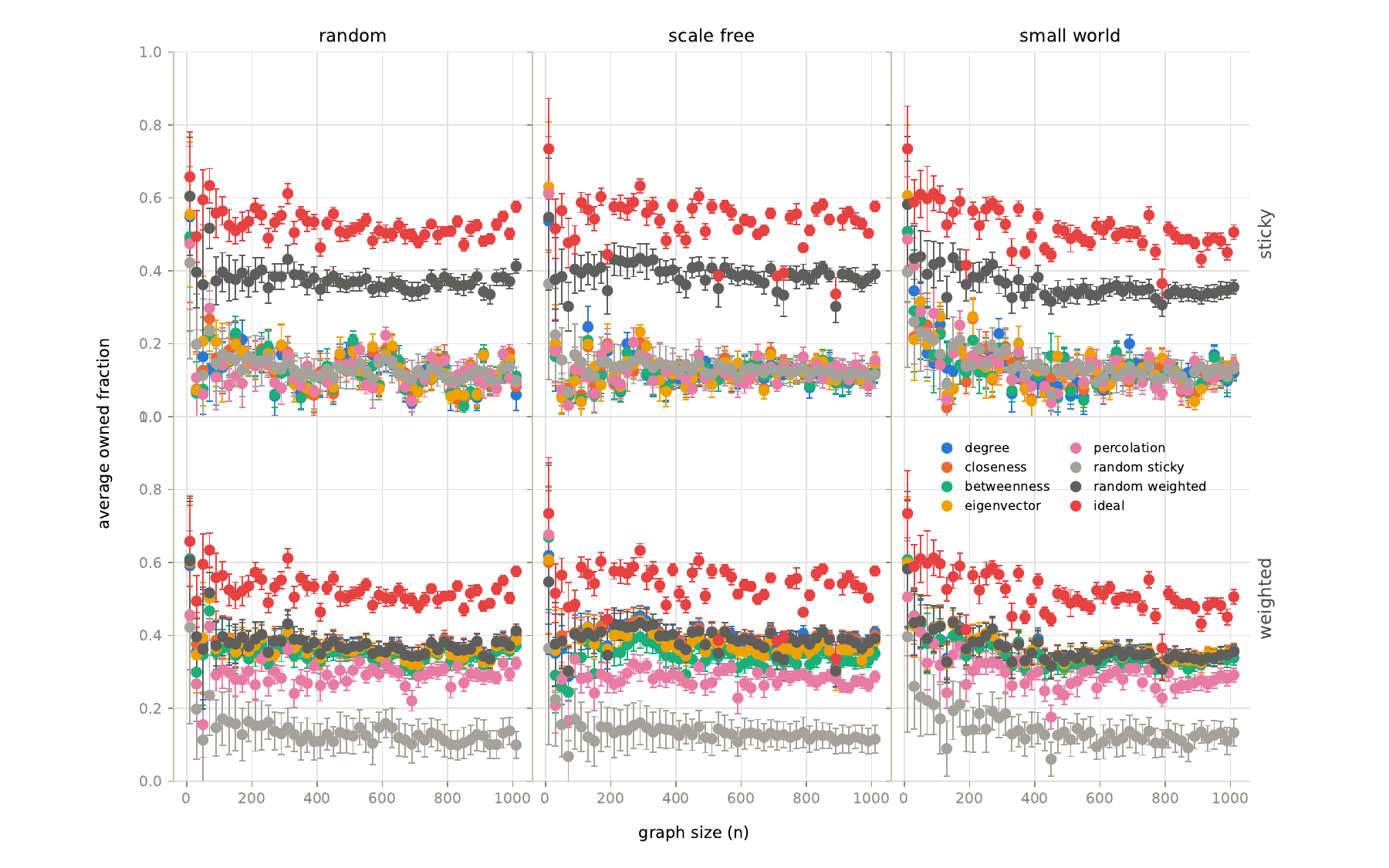}
     \caption{Percentage of owned hosts across increasing graph sizes $n$. The rows represent propagation strategies and columns graph topologies. Data points are means over 100 trials and bars represent the standard deviation.}
    \label{fig:graph_scaling}
\end{figure}
Our first experiments study the effects of attack scaling.
We highlight in Figure~\ref{fig:attack_scaling}
the average owned fraction of the graph
as the attacker budget grows.
Here, $k$ grows progressively with $1$ 
($k$ = $1$, $2$, .., $10$)
during the run, 
while the number of nodes $n$ and the number of edges $m$
stay the same.
There are $100$ trials with the same $k$, 
each with a different seed for computing the attack success probability.
We notice that,
as the attacker budget $k$ grows,
a hierarchy between the centralities starts to become visible.
Indeed,
we can see in Figure \ref{fig:attack_scaling} that
the important attack aspects are represented more by
the attacker's budget and strategy (sticky or weighted) 
than the actual victim choice.

We investigate next the effect of the network size.
In Figure~\ref{fig:graph_scaling}
the graph size (number of network nodes) 
varies with
$n = \{10, 30, \dots, 1010\}$
and
$m = 2.5 n$ edges
with a total of $51$ graphs.
The time horizon $t$ (or iterations) changes proportionally 
with $t$ = $3 \cdot n$.
As in the first experiment,
we have $100$ trials for every $n$.
The attacker budget remains constant at $k=1$.
A first remark regarding Figure~\ref{fig:graph_scaling} is that
the standard deviation of the average owned fraction of the network
shrinks as the graph size grows.
This makes sense,
as for smaller $n$ values
outliers are more visible
when the random attack or topology seeds favour one side
and can lead to a significant impact on the overall success of the botnet.
Larger $n$ values have these averaged away to some degree.
Thus the right side of each panel
is better at depicting the advantages of the chosen strategies across graph topologies
as the network size increases.

\begin{figure*}[t]
\subfloat[Random graph with $n = 1024$ nodes and $k = 8$ attacker budget.]{%
    \includegraphics[width=.94\textwidth]{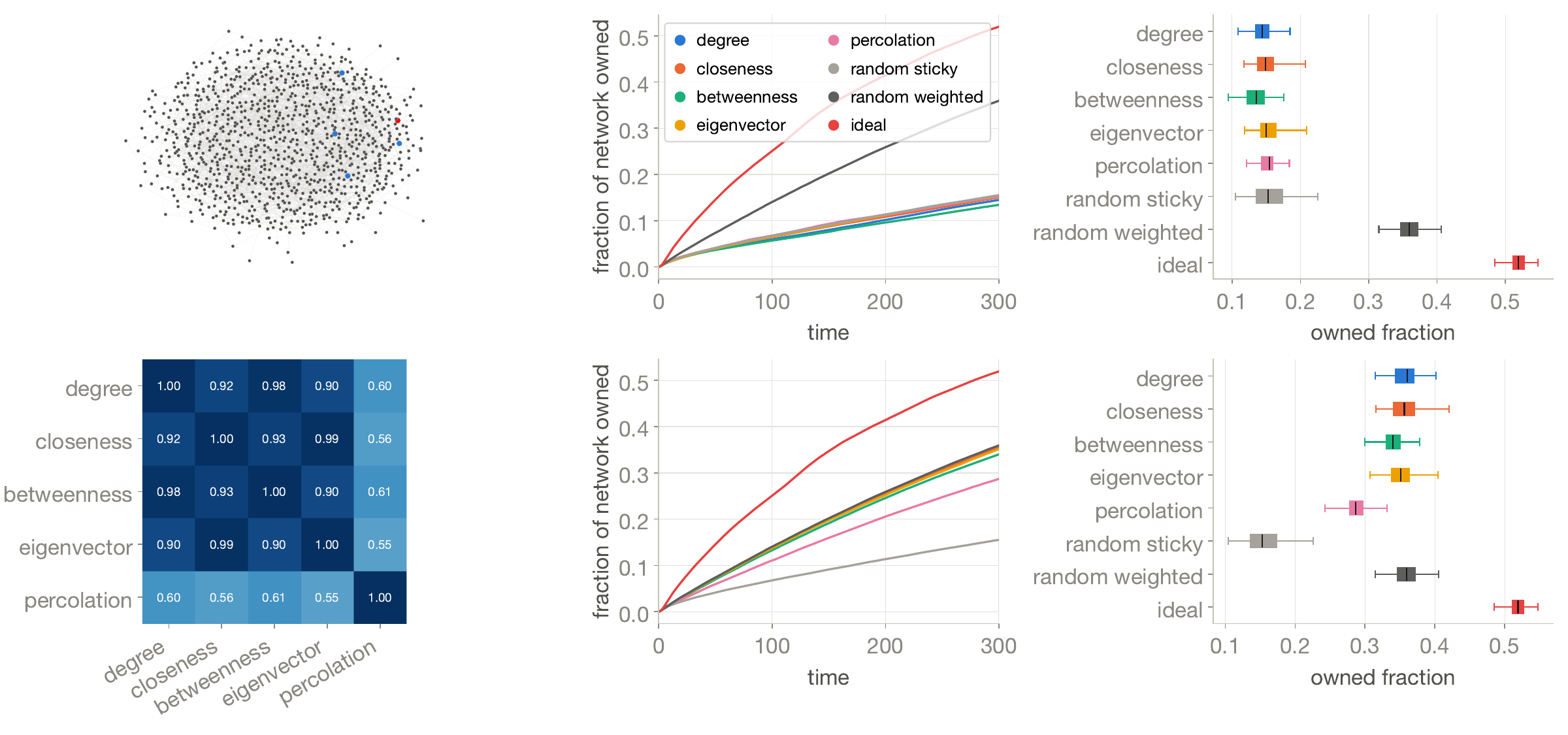}%
    \label{fig:random_1024_8}%
}\\
\subfloat[Scale-free graph with $n = 1024$ nodes and $k = 8$ attacker budget.]{%
    \includegraphics[width=.94\textwidth]{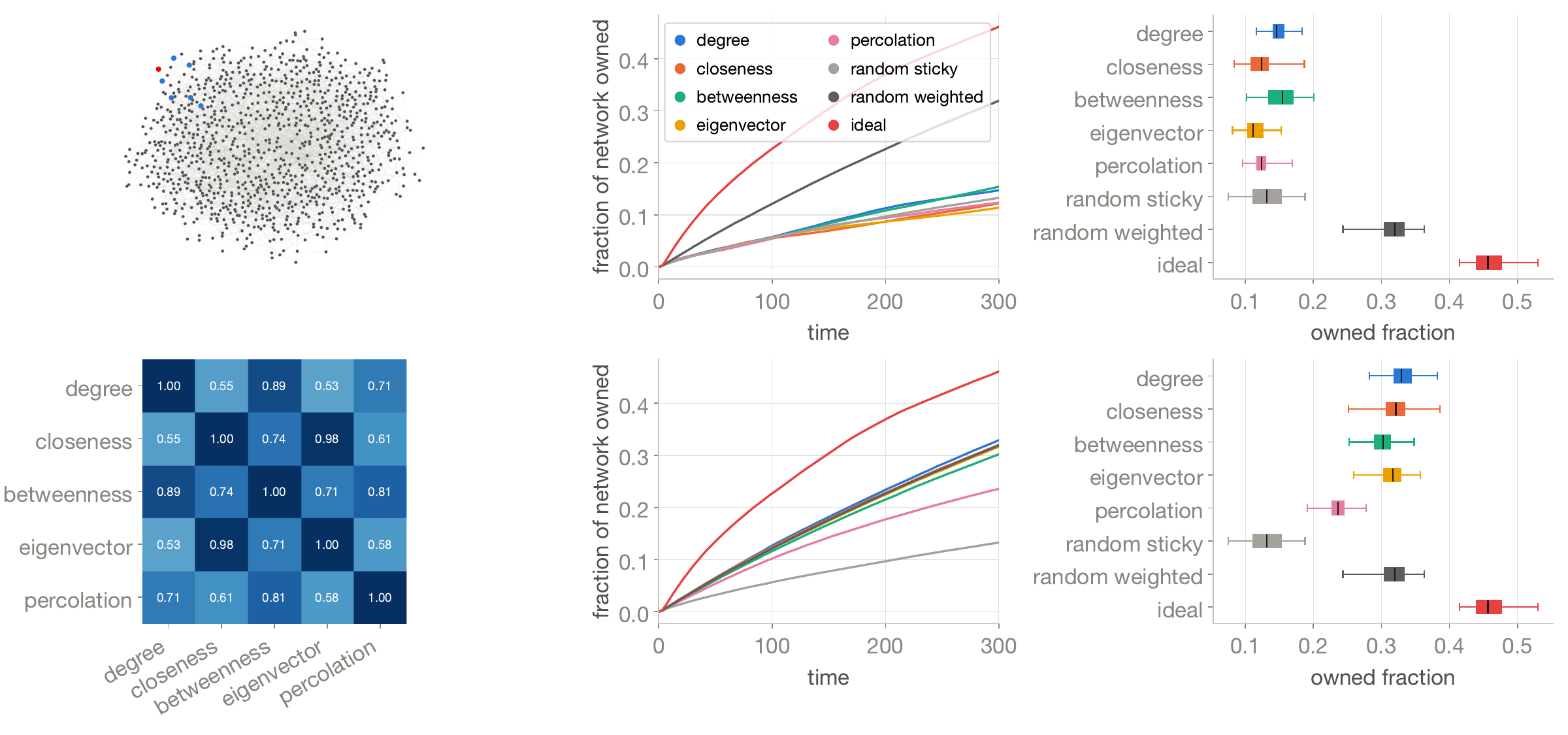}%
    \label{fig:scale-free_1024_8}%
}\\
\subfloat[Small-world graph with $n = 1024$ nodes and $k = 8$ attacker budget.]{%
    \includegraphics[width=.94\textwidth]{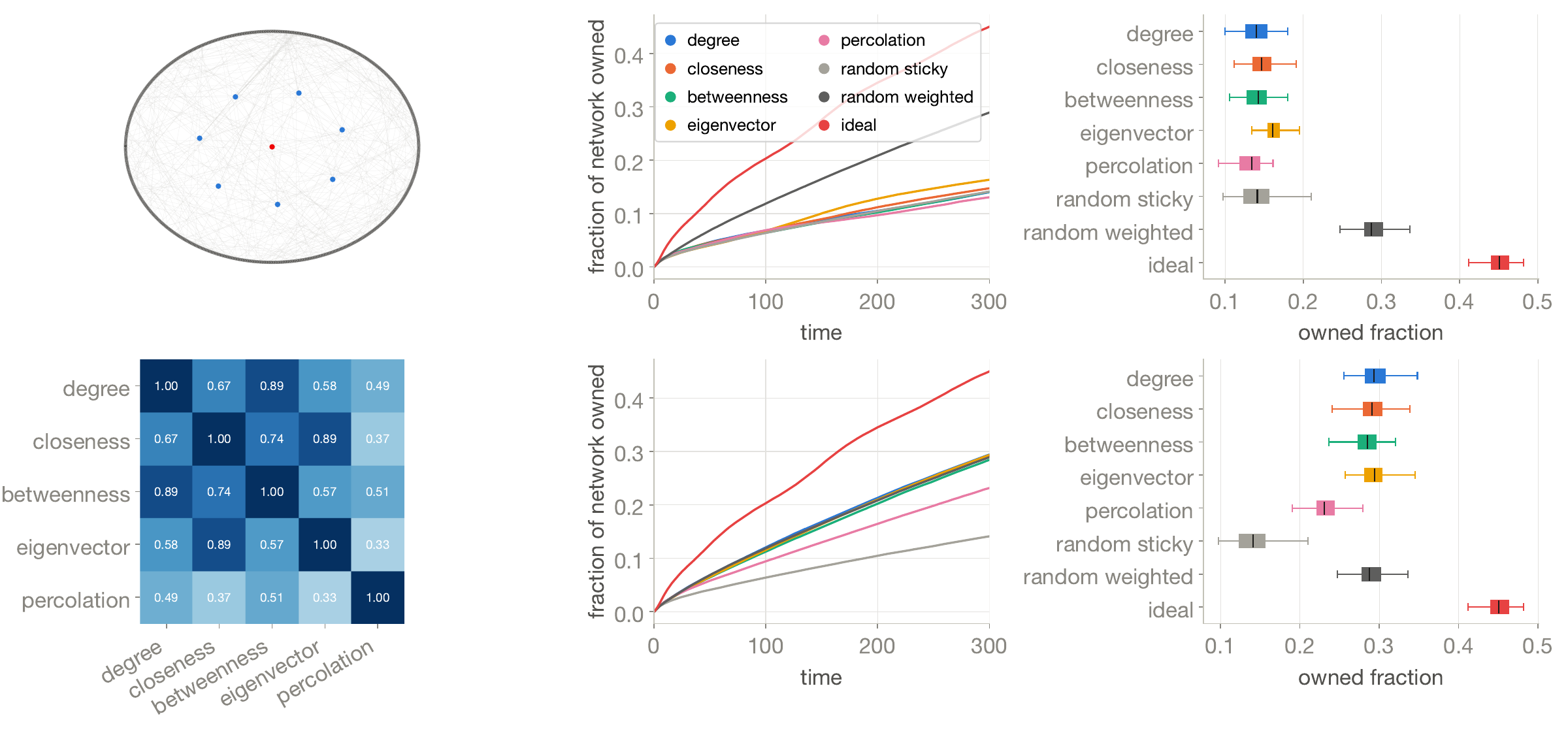}%
    \label{fig:small-world_1024_8}%
}
\caption{Spearman correlation study for graphs of 1024 nodes.}
\label{fig:static}
\end{figure*}
\begin{figure}[h!]
    \includegraphics[width=\textwidth]{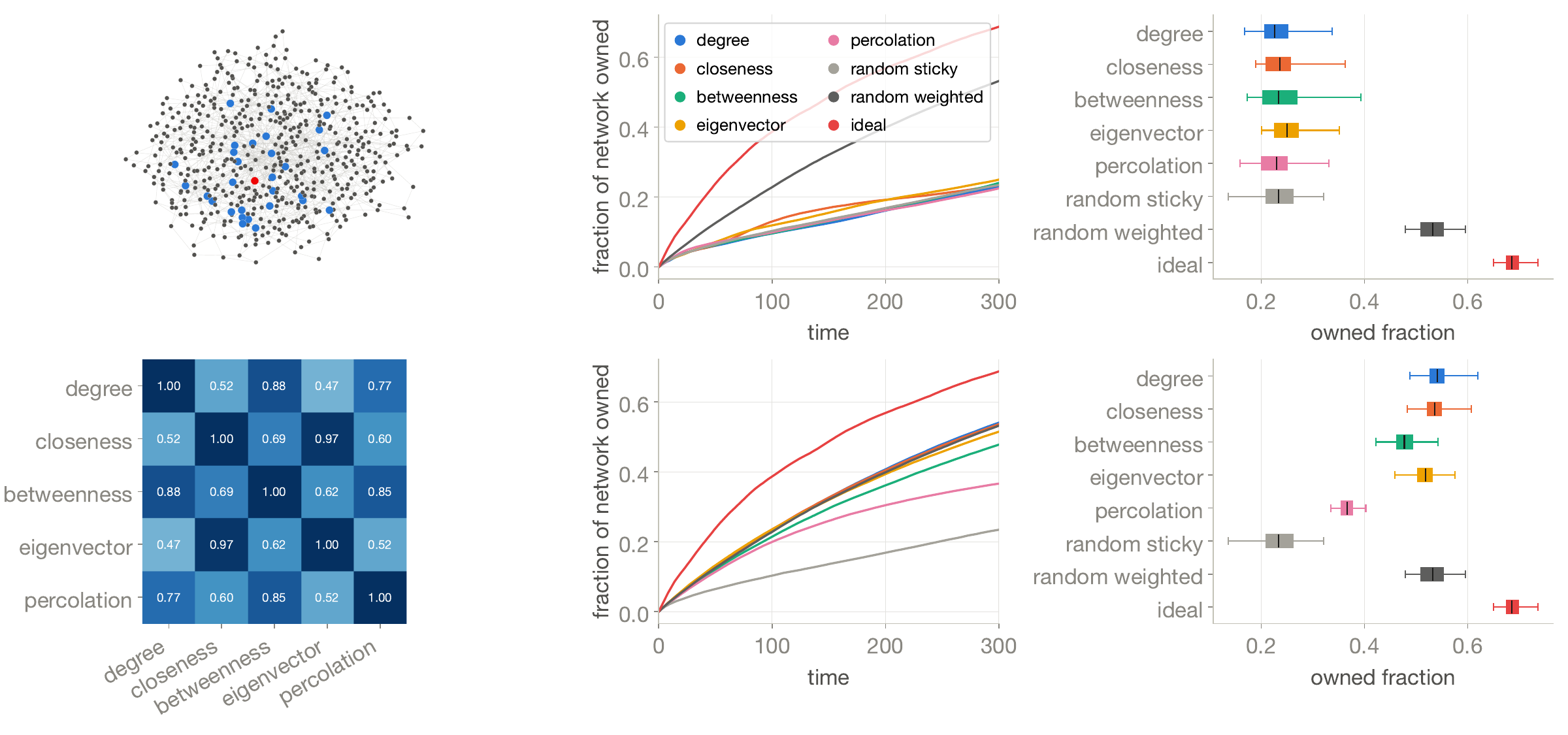}
    \caption{Scale-free graph with $n = 550$ nodes and $k = 8$ attacker budget.}
    \label{fig:scale-free_550_8}
\end{figure}

The experiments from Figure~\ref{fig:static}
feature runs with fixed number of
nodes $n$, edges $m$, attacker budget $k$ and iterations $t$.
Every panel reports
on the first column the network (with the botnet and its frontier marked)
and below the Spearman correlation matrix~\cite{Oldham-PONE-2019-Spearman}
of the centralities on the given topology;
on the second column contains the fraction of network owned across time instances;
while on the third column depicts the mean and standard deviation of owned fraction of each
strategy-centrality pairing over $100$ trials.
The top row shows the results of the sticky strategy
whereas the bottom one for the weighted ones.
Figures ~\ref{fig:random_1024_8}, ~\ref{fig:scale-free_1024_8}, ~\ref{fig:small-world_1024_8}
correspond to a random, scale-free and a small-world
network respectively,
each with $n=1024$, $m=300$ and $k=8$.
In the Appendix we also show
Figure~\ref{fig:static_550},
which follows the same order with $n=550$, $m = 1375$, $k=1$ and $t=300$,
and Figure~\ref{fig:scale-free_550_8} where we increase $k=8$.
%
Focusing on the structure of small world, we observe that
centralities remained grouped with their natural pairing.
One must look at the Figure~\ref{fig:graph_topologies} to observe that,
unless the attacker selects a shortcut,
at most $n_k$ additional nodes are added in $S^{(t)}$
where $n_k$ represents the neighbours of each host.
Figures \ref{fig:random_1024_8} and \ref{fig:random_550_1}
show that the Spearman correlation measurements of
\textit{degree}, \textit{closeness}, \textit{betweenness} and \textit{eigenvector}
have values larger than $0.9$.
This is expected as in random graphs there are no
hubs or clusters and every node is almost at the same distance
from other nodes.
Percolation is the odd one out, because under our formulation,
it counts only shortest paths originating from the attacker.
Precisely, it ranks nodes by their position in the BFS tree rooted
in $I^{(t)}$ rather than by their position in the graph.
Furthermore,
the average fraction of network owned by the attacker is comparable to their random
counterpart and we identify the only constant difference for
\textit{percolation} when utilising the \textit{weighted} strategy.
This remained true throughout our experiments and configurations,
suggesting that percolation, at least in the current form,
is less favourable for the attacker than random under any scenario.
Finally,
notice that in scale free networks the attacker's chance of
infiltrating the network is of importance.
The number of susceptible hosts in Figure~\ref{fig:scale-free_550_8}
at $t=0$ is significantly higher than in Figure~\ref{fig:scale-free_1024_8}
leading to a higher infection rate regardless of the chosen strategy.

\section{Conclusions and future directions}\label{sec:end}

In this paper we 
introduced the Host Attack Graph model
for estimating compromise probabilities of a given computer network
and proposed
two botnet propagation strategies.
A detailed study and Spearman analysis is provided
for random, scale-free and small-world graphs
where botnets choose their victims as they progress across time
based on
node degree, closeness, betweenness, eigenvector
and percolation centralities.
We also
introduced an ideal strategy
that we use for root greedy propagation
along with a random-choice strategy.
Our simulations show the effectiveness of our two strategies
and provide insights regarding the impact of the network topology on the botnet propagation.
Following reproducible research principles,
we make our code and data available online.

Future research,
as seen in the experiments,
should focus on centrality-based selection
as current methods provide modest improvements
over the baseline
regardless of network topology.
Of special interest is also the way success is measured during propagation,
while the number of bots in the network is important
we argue that not all bots are equal.
Indeed it might prove more important to target an isolated node
that contains valuable data (such as user credentials for example)
compared to a regular desktop.
In the future we
will look into
integrating centralities within the HAG model
in order to improve target selection.
We consider frontier $Pr(U(h))$'s to be of special interest here.
A few other interesting extensions to HAG are
host-budget per bot mimicking real-world distributed attacks
and
adapting percolation to dynamic time-dependent scenarios.


\subsubsection{\ackname}
This research is supported by the project “Romanian Hub for Artificial Intelligence - HRIA”, Smart Growth, Digitization and Financial Instruments Program, 2021-2027, MySMIS no. 351416.

\bibliography{bib}
\bibliographystyle{splncs04}
\section*{Appendix}

\subsection{Host Attack Graph Examples}

In this section we provide a few practical examples starting from the synthetic attack based on real vulnerabilities from Table~\ref{tab:vuln}
and the vulnerabilities chain in Figure~\ref{fig:hag}.

\begin{table}[h!]
\tabcolsep 5pt
\small
\begin{center}
\begin{tabular}{c | c c c c}
ID
& Name & Type & CVSS & $p_e(v)$\\
\hline
$v_1$ & CVE-2026-21510 & feature bypass  & 8.8 & 0.88\\
$v_2$ & CVE-2025-33053 & remote code execution & 8.8 & 0.88\\
$v_3$ & CVE-2025-24054 & spoofing & 6.5 & 0.65\\
$v_4$ & CVE-2026-32202 & spoofing & 4.3 & 0.43\\
$v_5$ & CVE-2025-21418 & privilege escalation &  7.8 & 0.78\\
$v_6$ & CVE-2025-24983 & privilege escalation  &  7.0 & 0.70\\
$v_7$ & CVE-2025-29824 & privilege escalation  & 7.8 & 0.78\\
\end{tabular}
\end{center}
\caption{Vulnerabilities for the example HAG in Figure~\ref{fig:hag}.
The table presents a synthetic example constructed using recent vulnerabilities found in Microsoft Windows.
It is not intended nor useful for use in an actual attack scenario.
It is included solely for demonstration purposes.}
\label{tab:vuln}
\end{table}

    \begin{table}[h!]
    \tabcolsep 5pt
    \small
    \begin{center}
    \begin{tabular}{c | c | c | c | c | c | c | c || c }
    $S_{v_1}$ & $S_{v_2}$ & $S_{v_3}$ & $S_{v_4}$ &
    $\bigwedge_{\pi_1}$ & $\bigwedge_{\pi_3}$ & $\bigwedge_{\pi_3}$ & $\bigwedge_{\pi_4}$ &
    $U(h)$ \\
    \hline
    0 & 0 & 0 & 0 &  
    0 & 0 & 0 & 0 &
    0 \\
    0 & 0 & 0 & 1 &  
    0 & 0 & 0 & 0 &
    0 \\
    0 & 0 & 1 & 0 &  
    0 & 0 & 0 & 0 &
    0 \\
    0 & 0 & 1 & 1 &  
    0 & 0 & 0 & 0 &
    0 \\
    0 & 1 & 0 & 0 &  
    0 & 0 & 0 & 0 &
    0 \\
    0 & 1 & 0 & 1 &  
    0 & 0 & 1 & 0 &
    1 \\
    0 & 1 & 1 & 0 &  
    0 & 0 & 0 & 0 &
    0 \\
    0 & 1 & 1 & 1 &  
    0 & 0 & 1 & 1 &
    1 \\
    1 & 0 & 0 & 0 &  
    0 & 0 & 0 & 0 &
    0 \\
    1 & 0 & 0 & 1 &  
    0 & 0 & 0 & 0 &
    0 \\
    1 & 0 & 1 & 0 &  
    0 & 0 & 0 & 0 &
    0 \\
    1 & 0 & 1 & 1 &  
    0 & 0 & 0 & 0 &
    0 \\
    1 & 1 & 0 & 0 &  
    1 & 0 & 0 & 0 &
    1 \\
    1 & 1 & 0 & 1 &  
    1 & 0 & 1 & 0 &
    1 \\
    1 & 1 & 1 & 0 &  
    1 & 1 & 0 & 0 &
    1 \\
    1 & 1 & 1 & 1 &  
    1 & 1 & 1 & 1 &
    1 \\
    \end{tabular}
    \end{center}
    \caption{Truth table for $U(h)$}
    \label{tab:Uh-truth-table}
    \end{table}
    
    \begin{table}[h!]
    \tabcolsep 5pt
    \small
    \begin{center}
    \begin{tabular}{c | c | c | c | c | c || c }
    $S_{v_5}$ & $S_{v_6}$ & $S_{v_7}$ &
    $\bigwedge_{\pi_5}$ & $\bigwedge_{\pi_6}$ & $\bigwedge_{\pi_7}$ &
    $R(h)$ \\
    \hline
    0 & 0 & 0 &  
    0 & 0 & 0 &
    0 \\
    0 & 0 & 1 &  
    0 & 0 & 0 &
    0 \\
    0 & 1 & 0 & 
    0 & 0 & 0 &
    0 \\
    0 & 1 & 1 &
    0 & 0 & 1 &
    1 \\
    1 & 0 & 0 &
    0 & 0 & 0 &
    0 \\
    1 & 0 & 1 &
    1 & 0 & 0 &
    1 \\
    1 & 1 & 0 &
    0 & 0 & 0 &
    0 \\
    1 & 1 & 1 &
    1 & 1 & 1 &
    1 \\
    \end{tabular}
    \end{center}
    \caption{Truth table for $R(h)$ assuming $U(h) = 1$}
    \label{tab:Rh-truth-table}
    \end{table}

\begin{example}
\label{ex:attack-paths}

    Let $h$ be a host that has the HAG given in Figure \ref{fig:hag}
    and the vulnerabilities of Table \ref{tab:vuln}.

    Observe that $v_2$ and $v_3$ become reachable by exploiting either $v_1$ or $ v_2$.
    The paths of HAG that grant access to $U(h)$ are:
    $\pi_1 = \{v_1, v_2\}$, $\pi_2 = \{v_1, v_2, v_3\}$,
    $\pi_3 = \{v_4, v_2\}$ and $\pi_4 = \{v_4, v_2, v_3\}$.
    Table~\ref{tab:Uh-truth-table} represents the truth table of the $\Pi(U(h))$ for $h$.

    One could observe that, in the given example,
    all attack paths are dependent
    on the state of $v_2$.
    This reflects real-world scenarios,
    in which an attacker may employ multiple attack vectors
    that depend on the exploitation of several vulnerabilities
    to gain unauthorised access.
    Therefore, the proposed HAG does not treat $\pi_1$ and $\pi_4$
    as independent events, since both are conditioned on $v_2$.

    Similarly, $\pi_5 = \{v_5, v_7\}$, $\pi_6 = \{v_5, v_6, v_7\}$,
    $\pi_7 = \{v_6, v_7\}$ are the attack paths available to gain
    $R(h)$ which are expressed in the truth table from Table~\ref{tab:Rh-truth-table}.
    
\end{example}

\begin{example}
    Based on the attack paths from the previous example,
    we can determine the probability of user compromise
\begin{align*}
        Pr(U(h)=1) &= Pr(\pi_1 = 1) + Pr(\pi_2 = 1) + Pr(\pi_3 = 1) + Pr(\pi_4 = 1) - {} \\
        &\quad - Pr(\pi_1 \cap \pi_2 = 1) - Pr(\pi_1 \cap \pi_3 = 1) - Pr(\pi_1 \cap \pi_4 = 1) - {} \\ 
        &\quad- Pr(\pi_2 \cap \pi_3 = 1) - Pr(\pi_2 \cap \pi_4 = 1) - Pr(\pi_3 \cap \pi_4 = 1) + {} \\ 
        &\quad+ Pr(\pi_1 \cap \pi_2 \cap \pi_3 = 1) + Pr(\pi_1 \cap \pi_2 \cap \pi_4 = 1) + {} \\
        &\quad+ Pr(\pi_1 \cap \pi_3 \cap \pi_4 = 1) + Pr(\pi_2 \cap \pi_3 \cap \pi_4 = 1) - {} \\
        &\quad- Pr(\pi_1 \cap \pi_2 \cap \pi_3 \cap \pi_4 = 1) \\
        &= p_e(v_1)p_e(v_2) + p_e(v_1)p_e(v_2)p_e(v_3) + p_e(v_2)p_e(v_4) + p_e(v_2)p_e(v_3)p_e(v_4) - \\
        &\quad- p_e(v_1)p_e(v_2)p_e(v_3) - p_e(v_1)p_e(v_2)p_e(v_4) - p_e(v_1)p_e(v_2)p_e(v_3)p_e(v_4) - \\
        &\quad- p_e(v_1)p_e(v_2)p_e(v_3)p_e(v_4) - p_e(v_1)p_e(v_2)p_e(v_3)p_e(v_4) -  p_e(v_2)p_e(v_3)p_e(v_4) +\\
        &\quad+ p_e(v_1)p_e(v_2)p_e(v_3)p_e(v_4) + p_e(v_1)p_e(v_2)p_e(v_3)p_e(v_4) + \\
        &\quad+ p_e(v_1)p_e(v_2)p_e(v_3)p_e(v_4) + p_e(v_1)p_e(v_2)p_e(v_3)p_e(v_4) - \\
        &\quad- p_e(v_1)p_e(v_2)p_e(v_3)p_e(v_4) \\
        &= p_e(v_1)p_e(v_2) + p_e(v_2)p_e(v_4) - p_e(v_1)p_e(v_2)p_e(v_4) \\
        &= 0.7744 + 0.3784 - 0.332992 = 0.8198
\end{align*}
\end{example}

\begin{example}
Based on the attack paths and the probability of user compromise from the previous examples,
we can finally determine the probability of root compromise
for the HAG in Figure~\ref{fig:hag}.
We start with the PPIE
\begin{align*}
    PPIE &= Pr(\pi_5=1) + Pr(\pi_6=1) + Pr(\pi_7=1) -\\
    &\quad - Pr(\pi_5 \cap \pi_6=1) - Pr(\pi_5 \cap \pi_7=1) - Pr(\pi_6 \cap \pi_7=1) + \\
    &\quad + Pr(\pi_5 \cap \pi_6 \cap \pi_7=1) =\\
    &= p_e(v_5)p_e(v_7) + p_e(v_5)p_e(v_6)p_e(v_7) + p_e(v_6)p_e(v_7) - \\
    &\quad - p_e(v_5)p_e(v_6)p_e(v_7) - p_e(v_5)p_e(v_6)p_e(v_7) - p_e(v_5)p_e(v_6)p_e(v_7) + \\
    &\quad + p_e(v_5)p_e(v_6)p_e(v_7) \\
    &= p_e(v_5)p_e(v_7) + p_e(v_6)p_e(v_7) - p_e(v_5)p_e(v_6)p_e(v_7) = \\
    &= 0.6084 + 0.546 - 0.42588 = 0.72852
\end{align*}
which results in $Pr(R(h)=1) = Pr(U(h)=1) \cdot PPIE
        = 0.8198 \cdot 0.72852
        = 0.5972 \approx 0.6$
\end{example}

\subsection{Spearman Experiments Continued}
In this section we complement the Spearman analysis of Section~\ref{sec:sim}
with simulations on smaller networks $n=550$.
We also summarize in Table~\ref{tab:centralities}
the centralities together with their characteristics and mathematical models
which should aid the reader in referencing these concepts.

\begin{table}[h!]
\tabcolsep 5pt
\renewcommand{\arraystretch}{2}
\small
\begin{center}
\begin{tabular}{p{1.8cm} | p{4.5cm} | c}
Centrality & Central node characteristics & Model\\
\hline
Degree & The number of distinct other nodes the node is connected to & $d_v = \frac{\deg(v)}{m}$\\
Closeness & How close the node is to other nodes (shortest paths) & $c_v = \dfrac{n-1}{\sum_{u=1}^{n-1} d(u,v)}$\\
Betweenness & How often the node is on the shortest path between other nodes & $b_v = \sum_{s,t \in V} \dfrac{\sigma(s,t \mid v)}{\sigma(s,t)}$\\
Eigenvector & Connected to other (influential or not) nodes & $e_v = \dfrac{1}{\lambda} \sum_{u \in N(v)} e_u$\\
Percolation & How often the node is on the shortest path between percolated and not percolated nodes
& $p_v = \frac{1}{n-2}\sum_{s \neq v \neq t} \frac{\sigma(s,t \mid v)}{\sigma(s,t)} \frac{x_s}{[\sum x_i]-x_v}$ \\ 
\end{tabular}
\end{center}
\caption{Centrality measures used in this paper and in the reviewed literature.}
\label{tab:centralities}
\end{table}

\begin{figure*}[h!]
\centering
\subfloat[Random graph with $n = 550$ nodes and $k = 1$ attacker budget.]{%
    \includegraphics[width=.93\textwidth]{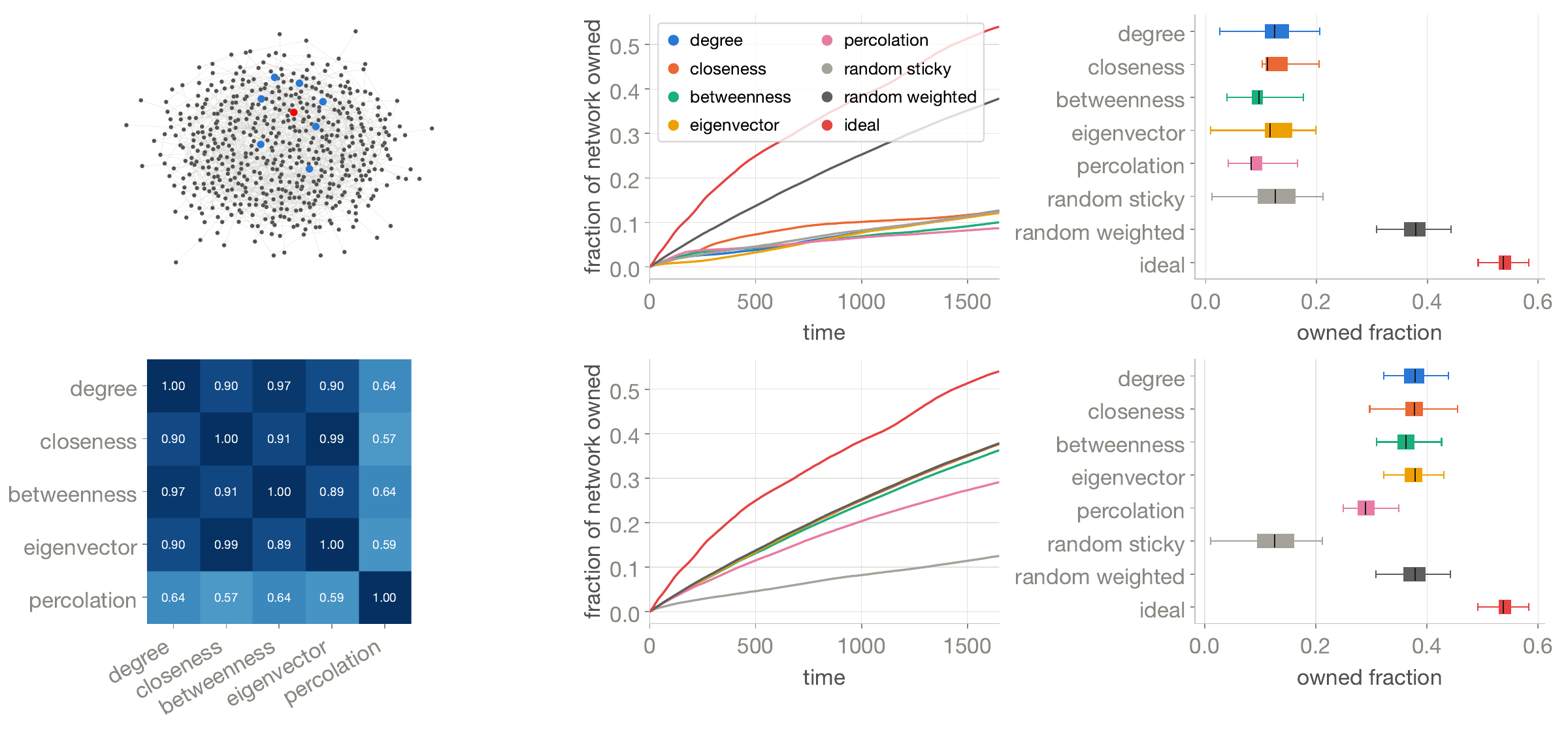}%
    \label{fig:random_550_1}%
}\\
\subfloat[Small-world graph with $n = 550$ nodes and $k = 1$ attacker budget.]{%
    \includegraphics[width=.93\textwidth]{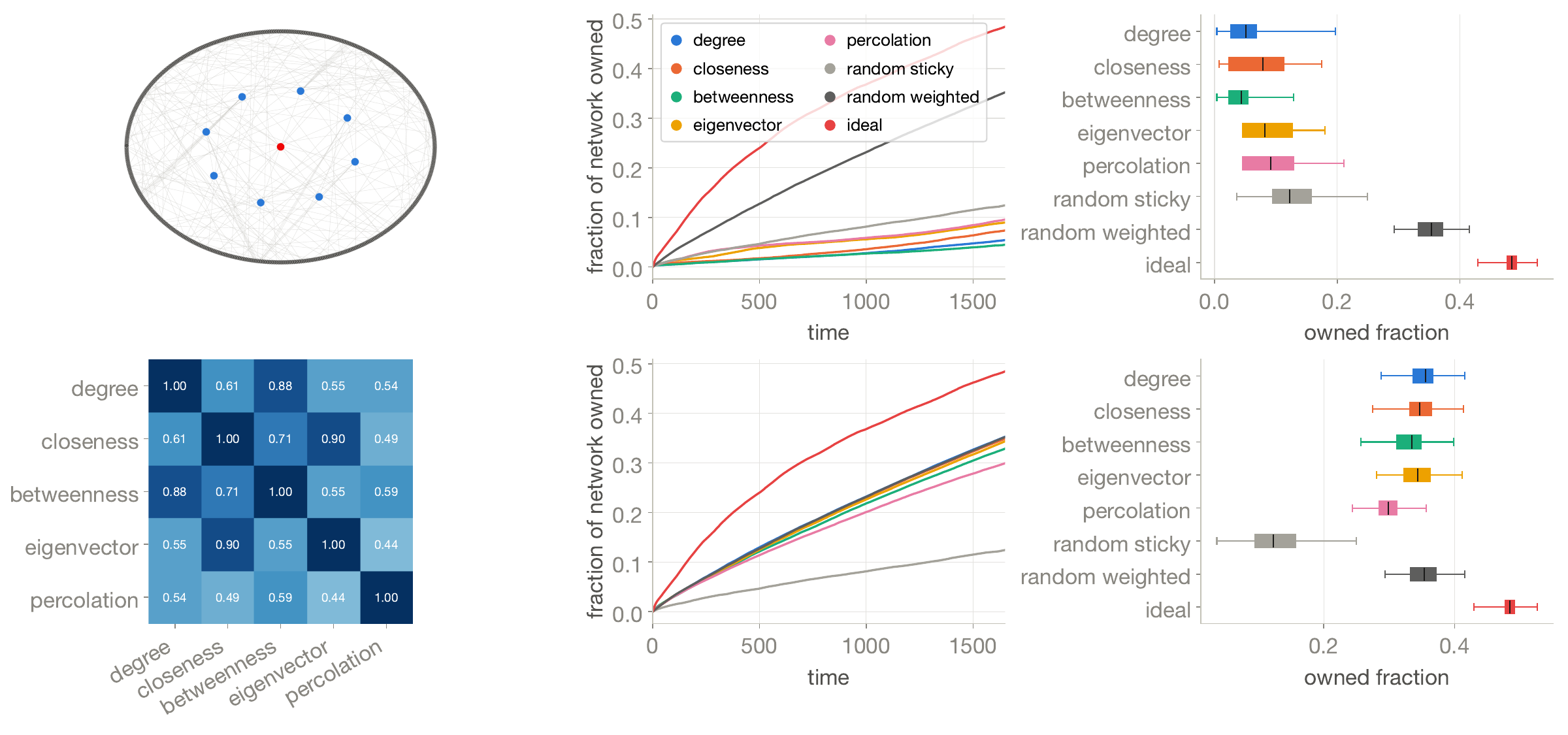}%
    \label{fig:small-world_550_1}%
}\\
\subfloat[Scale-free graph with $n = 550$ nodes and $k = 1$ attacker budget.]{%
    \includegraphics[width=.93\textwidth]{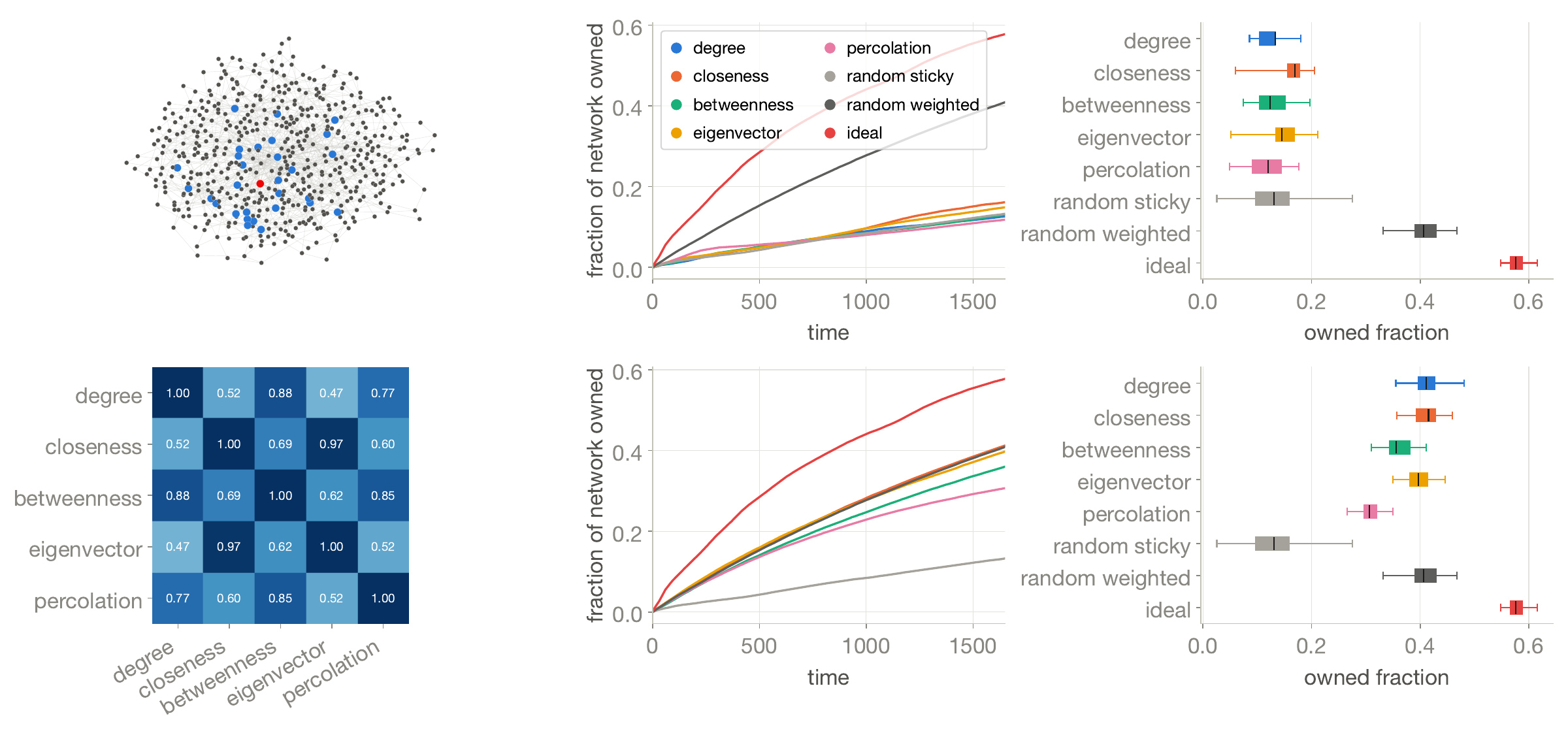}%
    \label{fig:scale-free_550_1}%
}
\caption{Spearman correlation study for graphs of 550 nodes.}
\label{fig:static_550}
\end{figure*} 
\end{document}